\documentclass[pdflatex,sn-mathphys-num]{sn-jnl}

\usepackage{graphicx}%
\usepackage{multirow}%
\usepackage{amsmath,amssymb,amsfonts}%
\usepackage{amsthm}%
\usepackage{mathrsfs}%
\usepackage[title]{appendix}%
\usepackage{xcolor}%
\usepackage{textcomp}%
\usepackage{manyfoot}%
\usepackage{booktabs}%
\usepackage{algorithm}%
\usepackage{algorithmicx}%
\usepackage{algpseudocode}%
\usepackage{listings}%

\theoremstyle{thmstyleone}%
\theoremstyle{thmstyletwo}%

\theoremstyle{thmstylethree}%

\AddToHook{shipout/firstpage}{%
    \raisebox{0.1cm}{%
     \makebox[\paperwidth][c]{%
  \hspace*{-0.28\paperwidth}%
  \footnotesize\bfseries
  Approved for public release; Distribution Unlimited. Public Release Case Number 26-0179}%
    }%
  }%

\begin{document}

\title[Article Title]{Foundry-Enabled Patterning of Diamond Quantum Microchiplets for Scalable Quantum Photonics}


\author*[1,2,3]{\fnm{Jawaher} \sur{Almutlaq}}\email{jawaher@mit.com}
\equalcont{These authors contributed equally to this work.}

\author[1]{\fnm{Alessandro} \sur{Buzzi}}
\equalcont{These authors contributed equally to this work.}

\author[4]{\fnm{Anders} \sur{Khaykin}}
\equalcont{These authors contributed equally to this work.}

\author[1]{\fnm{Linsen} \sur{Li}}

\author[4]{\fnm{William} \sur{Yzaguirre}}

\author[1]{\fnm{Maxim} \sur{Sirotin}}

\author[5]{\fnm{Gerald} \sur{Gilbert}}

\author*[4]{\fnm{Genevieve} \sur{Clark}}\email{gclark@mitre.org}

\author*[1,3]{\fnm{Dirk} \sur{Englund}}\email{englund@mit.edu}

\affil*[1]{\orgdiv{Research Laboratory of Electronics}, \orgname{Massachusetts Institute of Technology}, \orgaddress{\street{50 Vassar St}, \city{Cambridge}, \postcode{02139}, \state{MA}, \country{USA}}}

\affil[2]{\orgdiv{Materials Science and Engineering and Applied Physics}, \orgname{King Abdullah University of Science and Technology (KAUST)},  \city{Thuwal}, \postcode{23955}, \country{Saudi Arabia}}

\affil[3]{ \orgname{PhotonFoundries, Inc.}, 
\city{Brookline}, \postcode{02446}, \state{MA}, \country{USA}}

\affil[4]{ \orgname{The MITRE Corporation}, \city{Bedford}, \postcode{01730}, \state{MA}, \country{USA}}

\affil[5]{ \orgname{The MITRE Corporation}, \city{Princeton}, \postcode{08540}, \state{NJ}, \country{USA}}


\abstract{Quantum technologies promise secure communication networks and powerful new forms of information processing, but building these systems at scale remains a major challenge. Diamond is an especially attractive material for quantum devices because it can host atomic-scale defects that emit single photons and store quantum information with exceptional stability. However, fabricating the optical structures needed to control light in diamond typically relies on slow, bespoke processes that are difficult to scale. In this work, we introduce a manufacturing approach that brings diamond quantum photonics closer to industrial production. Instead of sequentially defining each device by lithography written directly on diamond, we fabricate high-precision silicon masks using commercial semiconductor foundries and transfer them onto diamond via microtransfer printing. These masks define large arrays of nanoscale optical structures, shifting the most demanding pattern-definition steps away from the diamond substrate, improving uniformity, yield, and throughput. Using this method, we demonstrate hundreds of diamond “quantum microchiplets” with improved optical performance and controlled interaction with quantum emitters. The chiplet format allows defective devices to be replaced and enables integration with existing photonic and electronic circuits. Our results show that high-quality diamond quantum devices can be produced using scalable, foundry-compatible techniques. This approach provides a practical pathway toward large-scale quantum photonic systems and hybrid quantum-classical technologies built on established semiconductor manufacturing infrastructure.
}




\maketitle

\section{Introduction}\label{sec1}

Color centers in diamond are leading solid-state qubits for quantum networks and quantum information processing due to their long spin coherence times, optical addressability, and ability to generate spin--photon entangled states in a wide-bandgap, high-index host material \cite{Aharonovich2016,Atature2018,Doherty2013,Loncar2013}. Embedding these emitters in diamond nanophotonic structures, such as waveguides and optical cavities, enhances photon extraction efficiency and enables deterministic spin--photon interfaces, which are essential for scalable quantum photonic architectures \cite{Schroder2016,Burek2014,Mouradian2015}.

Substantial progress has been made in fabricating high-quality diamond nanophotonic devices. Early demonstrations achieved one- and two-dimensional photonic crystal cavities in single-crystal diamond with quality factors exceeding \(10^{4}\)--\(10^{5}\) using electron-beam lithography and plasma etching \cite{Hausmann2010,RiedrichMoller2012,Burek2014,Lee2014}. Subsequent work improved performance and yield through optimized etch processes and hard-mask strategies \cite{Li2015,Khanaliloo2015,RiedrichMoller2016Nanoscale}. More recently, chiplet-based approaches have enabled heterogeneous integration of diamond nanophotonic devices with photonic integrated circuits (PICs) and CMOS platforms using pick-and-place and transfer-based techniques \cite{Hausmann2012,Kim2017,Wan2020,Elshaari2020}.

Despite these advances, scalable patterning of diamond nanophotonic structures remains challenging. Most state-of-the-art methods rely on direct resist patterning and electron-beam lithography on diamond, limiting throughput, reproducibility, and scalability to large arrays \cite{Hausmann2010,RiedrichMoller2012,RiedrichMoller2016Nanoscale}. Wafer-scale approaches based on thin-film diamond and related wide-bandgap materials offer dense integration but require uniform large-area membranes and provide limited post-fabrication flexibility \cite{RiedrichMoller2016Nanoscale,Ding2024NatCommun,Lukin2020}. Chiplet-based integration alleviates several of these constraints by enabling modular assembly and post-fabrication selection while preserving the advantages of single-crystal diamond \cite{Hausmann2012,Kim2017,Elshaari2020}. However, prior demonstrations of heterogeneous integration using legacy silicon nitride (SiN) hard-mask processes have been limited to cavity quality factors in the range of $\sim 600$-$900$ in photonic integrated circuit implementations \cite{Chen2024OpticaQ}, motivating improved fabrication approaches that are compatible with scalable, foundry-based manufacturing.

Here, we introduce a scalability pathway based on foundry-fabricated silicon (Si) hard masks for patterning single-crystal diamond quantum microchiplets. By transferring high-resolution pattern definition to a commercial semiconductor foundry, this approach enables parallel and reproducible fabrication without direct lithography on diamond with cavity quality factors improved by approximately $3$--$8\times$ relative to prior fabrication and heterogeneous integration demonstrations.
 The resulting chiplet-based platform preserves post-fabrication selection and compatibility with established photonic integrated circuit (PIC) and CMOS integration schemes, providing a practical route toward large-scale integrated quantum photonic systems \cite{Chen2024OpticaQ,Clark2024NanoL,Riedel2025,Harris2025arXiv}.

\begin{figure*}[b]
    \centering
    \includegraphics[scale=0.33]{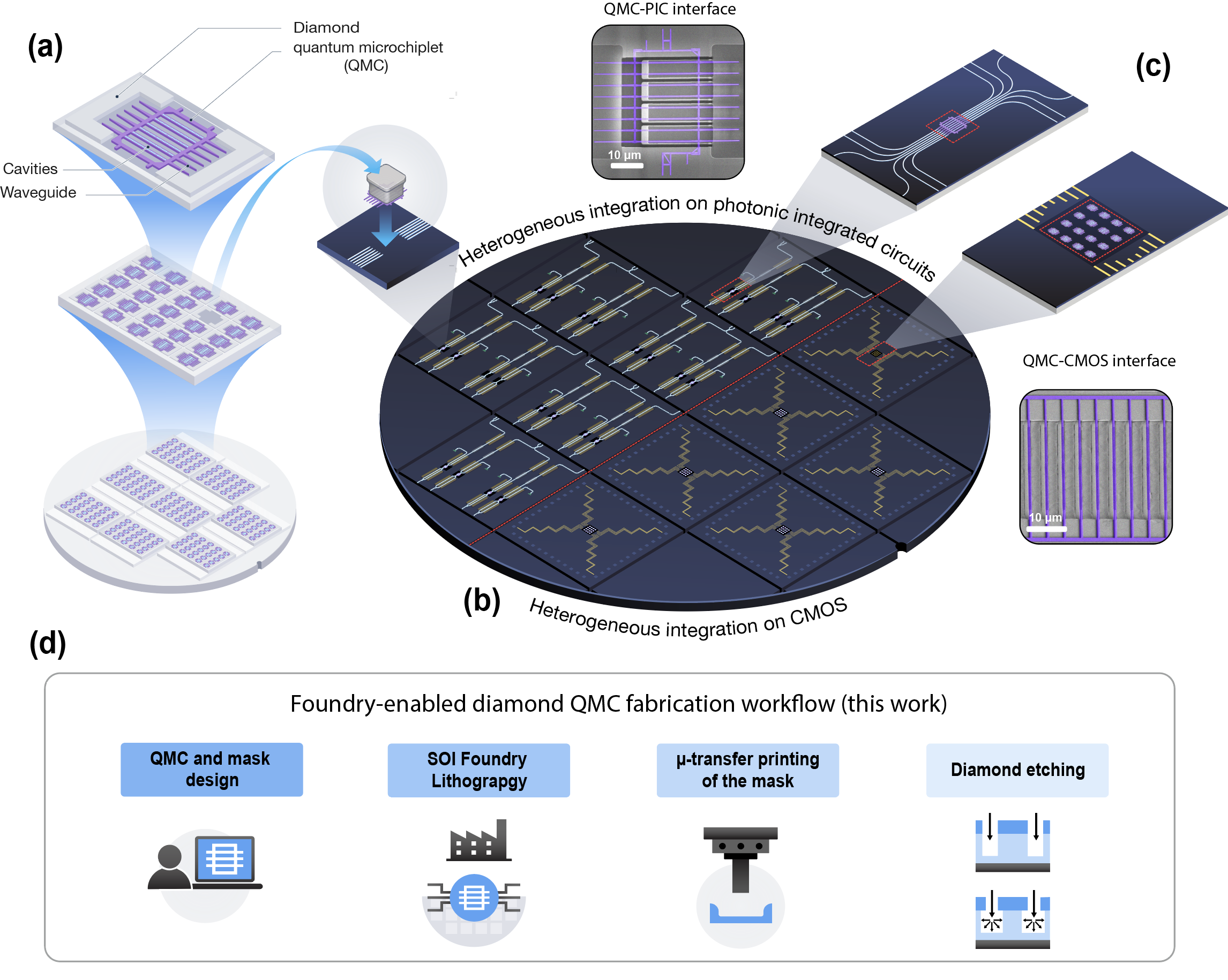} 
    \caption{Foundry-enabled diamond QMC fabrication workflow. \textbf{a)} Schematic of a diamond QMC, comprising nanophotonic waveguides and cavities patterned in single-crystal diamond using a transfer-printed silicon (Si) hard mask. Arrays of QMC chiplets are fabricated at the wafer scale. \textbf{b)} Wafer-scale heterogeneous integration of diamond QMCs onto CMOS substrates, enabling dense co-integration of diamond nanophotonics with electronic circuitry. \textbf{c)} Heterogeneous integration of diamond QMCs on photonic integrated circuits (PICs), illustrated with representative layouts and microscope/SEM images showing photonic routing and cavity arrays. \textbf{d)} Fabrication workflow: QMC and mask design, foundry mask fabrication, $\mu$-transfer printing of the Si hard mask onto diamond, and diamond etching to define QMCs. This work addresses the foundry-enabled silicon mask fabrication and diamond patterning steps highlighted here; heterogeneous integration with PIC and CMOS platforms is shown as an enabled capability.}
    \label{fig:CMOS_platform}
\end{figure*}

Relative to prior membrane-transfer and chiplet-based approaches, this work demonstrates: (i) wafer-scale enabled fabrication of silicon hard masks using commercial foundry lithography; (ii) high-yield transfer of large-area suspended membranes via commercial micro-transfer printing; and (iii) statistically uniform arrays of improved-Q diamond nanophotonic cavities with verified coupling to solid-state quantum emitters.

\section{Concept and Design}\label{sec2}

While electron-beam lithography (EBL) is used to fabricate the silicon hard mask, its role here is fundamentally different from direct EBL on diamond. In this approach, EBL is performed on a foundry-compatible silicon-on-insulator wafer to define a pattern carrier, after which all diamond nanostructures are fabricated without any direct lithography on the quantum material, enabling wafer-scale parallelism and improved reproducibility.

Silicon hard-mask transfer has been demonstrated previously, including via tungsten-tip-assisted PDMS transfer \cite{Li2015}; however, those implementations were limited to masks of order $\sim$200~$\mu$m and relied on manual, low-throughput handling. The present work introduces a scalable membrane geometry and commercial micro-transfer printing to enable reliable transfer of large-area ($750~\mu$m $\times$ $750~\mu$m) silicon masks containing dense arrays of QMCs.

The design consists of photonic crystal cavities embedded in 300-nm-wide nanobeam waveguides, patterned with 127-nm air holes optimized for coupling to Sn-117 color centers. Through engineering the membrane layout, we can suspend delicate waveguide regions while maintaining a robust outer frame, enabling large-area patterning and high-yield transfer. This modular design not only facilitates scalable fabrication across many devices but also ensures compatibility with standard semiconductor processing flows.

Each transferred membrane in this demonstration contains an array of 8 rows by 15 columns of QMCs, with each chiplet comprising 15 photonic waveguides. This membrane-based architecture enables wafer-scale fabrication and integration with sub-linear scaling. Let $E$ denote the total integration effort and $N$ the number of integrated QMCs. In a batch transfer-printing process, the integration effort can be expressed as
$E(B) = E_0 + B E_u$,
where $E_0$ represents the fixed overhead associated with a transfer-printing event (including handling, alignment, and setup), $E_u$ is the incremental effort per transferred membrane, and $B$ is the number of membranes transferred. Each membrane contains $N_c$ QMCs such that $N = B N_c$. The integration effort per chiplet is therefore given by
$\frac{E}{N} = \frac{E_0}{B N_c} + \frac{E_u}{N_c}$. As the number of transferred membranes increases, the integration effort per chiplet approaches a constant value set by the incremental transfer cost, demonstrating sub-linear scaling of integration effort with chiplet count. In a direct-write fabrication flow based on electron-beam lithography on diamond,
lithography must be repeated for each diamond chip, leading to fabrication time
and cost that increase in direct proportion to the number of patterned devices.
This effect is further amplified by the use of small, electrically insulating
diamond substrates with high charging sensitivity, which require
substrate- and operator-dependent tuning. This amortized-effort scaling establishes the basis for wafer-scale patterning and integration of large arrays of diamond QMCs.

\section{Fabrication of Diamond Quantum Microchiplets}\label{sec3}

\subsection{Si Hard Mask Fabrication and Suspension}\label{subsec3.1}

The silicon hard mask was fabricated on a commercial SOI wafer consisting of a 220~nm Si device layer and a 1~$\mu$m buried oxide. Photonic crystal nanobeam cavities were patterned into the Si device layer using electron-beam lithography, followed by reactive ion etching to transfer the design. A square membrane window (750~$\mu$m $\times$ 750~$\mu$m) was then defined, supported by an array of tethers (400~nm wide, 10~$\mu$m long, 31 per side) that mechanically anchored the membrane to the surrounding Si frame. To release the membrane, the underlying buried oxide was removed by immersion in 49\% hydrofluoric acid, leaving the patterned Si layer suspended by the tethers. A final critical point drying step was performed to avoid stiction and collapse of the thin membrane, resulting in a mechanically stable, free-standing hard mask ready for transfer. While the devices demonstrated here were released using wet HF etching followed by critical point drying, we note that for scalable manufacturing and foundry-compatible workflows, vapor-phase HF release provides a more robust and scalable alternative. Future implementations will leverage vapor HF to enable higher throughput and improved yield.

\subsection{Micro-Transfer Printing}\label{subsec3.2}
The silicon hard-mask membrane (750~$\mu$m $\times$ 750~$\mu$m, 220~nm thick) was released from an SOI wafer that had been patterned and under-etched, leaving the membrane suspended by the tethers.
Micro-transfer printing was performed in the MIT.nano class-100 cleanroom with an X-Display\,MTP-1002 tabletop printer.
The source chip holding the suspended mask and the target diamond chip were each mounted on silicon carrier wafers with copper tape and aligned on the printer stage.
A commercial PDMS stamp from X-Celeprint (600~$\mu$m $\times$ 600~$\mu$m), fixed to a glass slide, was lowered to contact the membrane; retraction of the stamp mechanically fractured the tethers, allowing the mask to be lifted.
The stamp carrying the mask was aligned to the diamond chip and brought into contact.
During retraction, a lateral shear of a few micrometres facilitated release of the mask onto the diamond.
Although the present transfer did not demand high overlay accuracy, the same procedure can achieve sub-micron alignment precision. The full diamond fabrication flow following mask transfer is described in the Methods and summarized in Extended Data Fig. 1.

\begin{figure*}[t]
    \centering
    \includegraphics[scale=0.65]{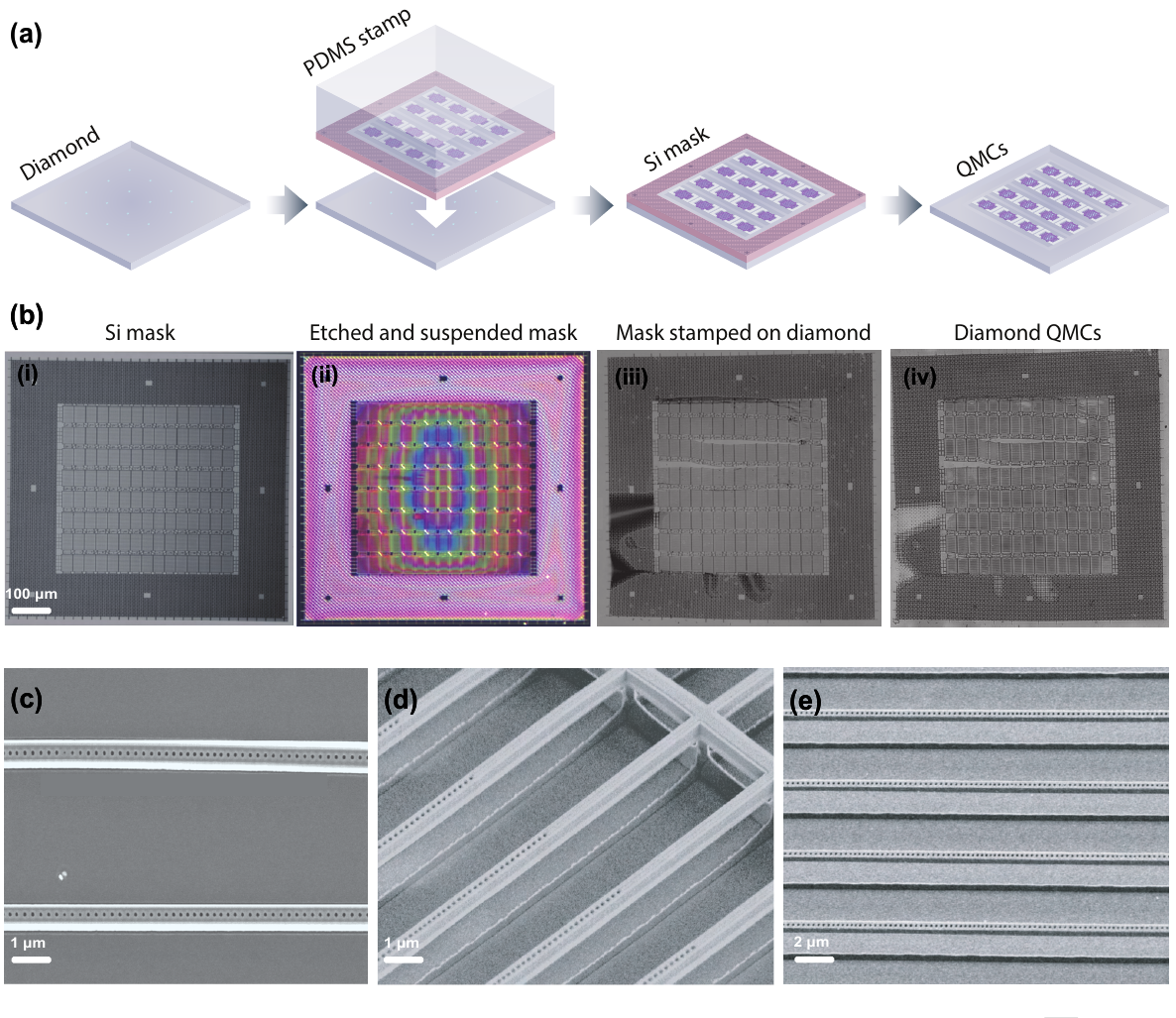} 
    \caption{Schematic illustration of the fabrication process. \textbf{a)} Silicon hard masks patterned on commercial silicon-on-insulator (SOI) wafers using foundry lithography are released and transferred onto bulk diamond substrates via micro-transfer printing. The transferred mask defines arrays of nanobeam cavities and waveguides during subsequent oxygen plasma etching, followed by mask removal to yield suspended diamond nanophotonic structures.
\textbf{b)} Optical micrographs of the fabrication process: (i) silicon mask prior to release, (ii) etched and suspended silicon membrane mask (shown in white-light illumination highlighting interference effects), (iii) mask stamped on diamond, and (iv) fabricated diamond QMCs.
\textbf{(c-e)} Scanning electron microscope (SEM) images of the resulting diamond nanophotonic structures, highlighting waveguides and cavity regions patterned with high fidelity.}
    \label{fig:Fabrication}
\end{figure*}

\section{Optical Characterization}\label{sec5}
\subsection{Emitter characterization and cavity measurements}\label{subsec5.1}

Figure 3a shows an optical microscope image of an 8 $\times$ 15 array of chiplets containing photonic crystal nanobeam cavities fabricated using our Si-hardmask method. The diamond substrate was previously implanted with the 117 tin isotope and annealed to create negatively charged tin vacancy centers. We investigate the optical quality of the nanostructures using photoluminescence measurements at cryogenic and room temperatures. Confocal raster scans of an exemplary chiplet show a uniform, bright optical response throughout the chiplet (Fig. 3b) while PL spectra measured at a location on a nanobeam away from any cavity structures show peaks at 619 and 620 nm from the C and D transition of tin vacancy centers created in the diamond. 

We next use hyperspectral raster scans to systematically characterize the optical properties of the photonic crystal cavities through coupled PL from SnVs in the nanostructure. A peak finding algorithm identifies cavity modes based on prominence and linewidth, and indexes them by location, linewidth, and resonance wavelength. Figure 3di shows a black and white raster scan of integrated PL intensity with the overlaid results of our peak finding showing the center wavelength of cavities in the chiplet. PL scans at the location of a detected cavity mode (Fig. 3dii, dark purple) show a pronounced resonance near 619.5 nm compared to spectra away from the cavity mode (light purple curve). Histograms of resonance wavelength and $Q$ value for cavities identified in this chiplet reveal a narrow distribution centered at 620 nm with $Q$ values ranging from $\sim$2500 to near 5000 at room temperature. 

We extend this systematic characterization to all chiplets in the mask gaining statistics on resonance wavelength, $Q$, and cavity yield to inform fabrication improvements. Histograms of cavity resonance wavelength and $Q$ value reveal a distribution of resonances centered near 635nm (Fig. 3ei), slightly lower energy than the target wavelength of 620 nm, with the distribution of $Q$ values centered near 3000 (Fig. 3eii). We quantify the yield of successful cavities by calculating the "fill-factor" for each chiplet in the mask, as the fraction of nanobeams in the chiplet containing at least one cavity resonance. We find that a majority of chiplets with cavities present have more than half of the nanobeams with cavities successfully resolved. To gain further insight into these parameters, Figure 3e iv and v show spatially resolved maps of resonance wavelength and $Q$ as well as their standard deviations (<s>) for each chiplet in the mask. Regions that appear white correspond to chiplets that did not show any cavity resonances. We distinguish between (i) \emph{structural yield}, determined by the mechanical integrity of the transferred mask, and (ii) \emph{cavity yield}, quantified by the fill fraction among structurally intact chiplets.
Comparing these spatial maps with the optical microscope image in panel 3a), there is a correspondence between chiplets with no cavity resonances and regions where the Si hardmask sustained damage or distortion during transfer. Considering only regions where the mask is visually intact, the cavity yield is substantially higher, indicating that yield can be improved through more careful mask transfer.

\begin{figure*}[!t]
    \centering
    \includegraphics[scale=0.55]{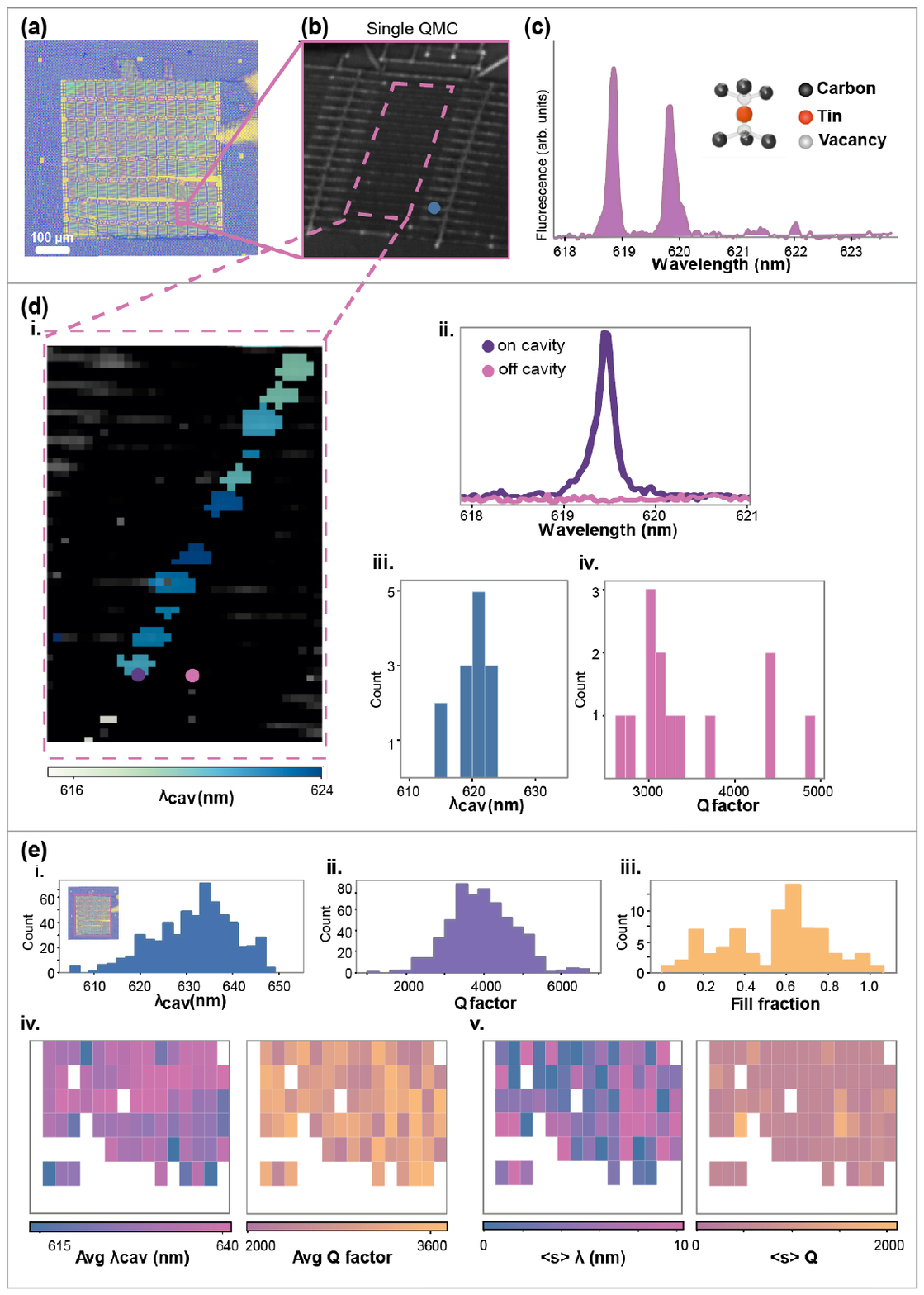}
    \caption{Optical characterization and cavity measurements. \textbf{a)} Optical microscope image of fabricated nanostructures. \textbf{b)} Raster scan of a single chiplet showing summed confocal photoluminescence (PL) at each location. \textbf{c)} PL spectrum from the location marked by the blue dot in b). \textbf{d)} Cavity characterization for a single chiplet. i) Hyperspectral map of the region in the dashed box in b. Cavities are identified using a peak finding algorithm and colored according to their center frequency. ii) Example spectrum on (dark blue) and off (light blue) a cavity. iii) and iv) histograms showing distribution of center frequency and $Q$ for this chiplet. \textbf{e)} Cavity characterization across the entire mask for all 120 chiplets. i), ii),iii) histograms showing center frequency and $Q$ distribution, as well as fill fraction. iv) spatial maps of center frequency and $Q$ with standard deviations for each (<s>) shown in v).  }
    \label{fig:OpticalCharacterization}
\end{figure*}

\subsection{Gas tuning}\label{subsec5.2}

Practical systems for quantum computing and networking require scaling to many cavity-coupled spin-photon interfaces, all operating at a single controllable wavelength. While fabrication imperfections inevitably cause some degree of inhomogeneity in cavity wavelength, we employ gas tuning to controllably tune the resonance wavelength of our cavities and demonstrate controllable coupling between cavities and color centers in our sample. We use cross-polarized reflection measurements to monitor cavity resonance shifts with gas adsorption, since reflection measurements do not require high excitation powers that can cause desorption of the gas from the nanobeams. Figure 4a shows the cross-polarized reflection spectrum from a cavity with a clear Fano-Lorentz resonance near 625 nm. Overlaid and fitted reflection and PL spectra show close agreement between measured resonance wavelengths and similar $Q$ values ($\sim 850$ for reflection, $1030$ for PL). Figure 4b shows the shift in resonance wavelength for a different cavity with a resonance near 640 nm as nitrogen gas is condensed on the sample. We measure similar shifts in resonance as a function of gas adsorption for cavities throughout the sample, with up to $\sim 10$ nm tuning range, enabling post-fabrication spectral alignment of cavity modes (see Extended Data Fig. 5). Based on the standard deviation we measure for cavity resonances within a chiplet (Fig. 3e), this tuning range should allow alignment of all cavities within high performing chiplets given our present fabrication tolerances in this sample.

After measuring a reproducible cavity resonance shift in cross-polarized reflection, we switch to PL for the same 640 nm cavity and monitor the tin emitter resonance at 645 nm as the cavity red shifts through the 645nm transition. We see a clear enhancement of roughly a factor of four when the cavity is in resonance with the transition at 645 nm. Taken together, these measurements allow us to quantify yield at each stage of the fabrication and optical characterization pipeline, summarized in Table~\ref{tab:yield_metrics}.

\begin{figure*}[htb]
    \centering
    \includegraphics[scale=0.41]{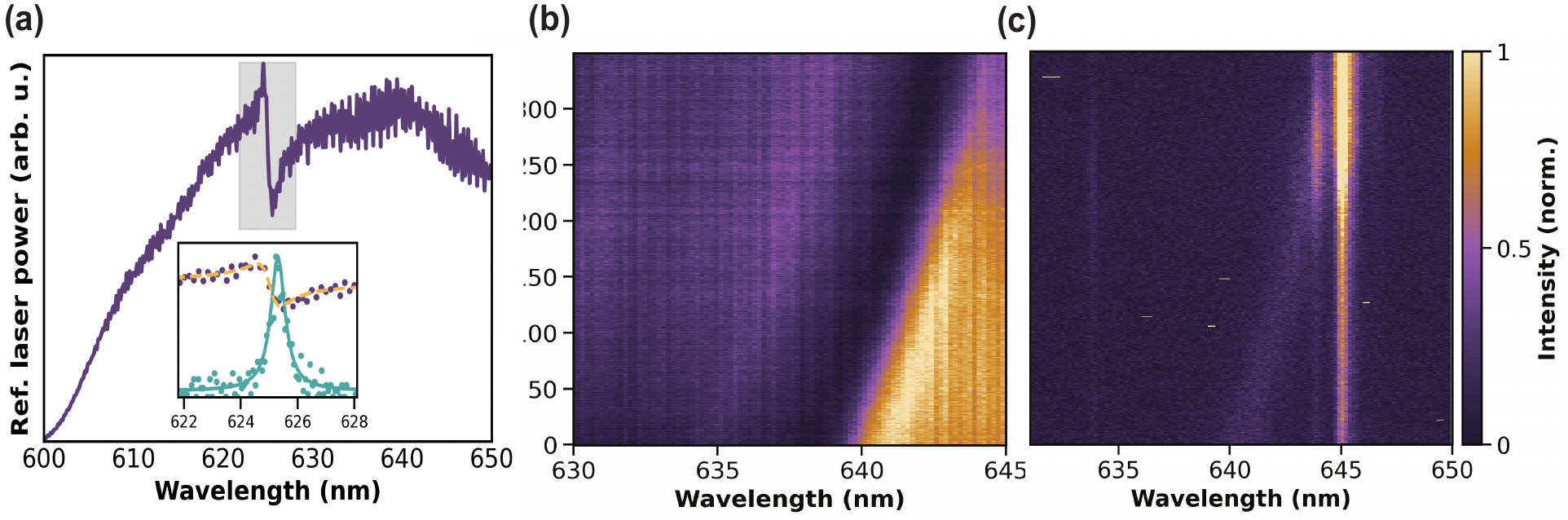} 
    \caption{Representative gas tuning of cavity resonances and Purcell enhancement of coupled SnVs. \textbf{a)} Cavity reflection spectrum showing a Fano resonance near 625 nm. i) Overlaid reflection and PL spectra for the cavity in a). \textbf{b)} Reflection spectrum as a function of time as nitrogen gas is added to the chamber. \textbf{c)} PL spectra showing the 645nm SnV peak as a function of time as the cavity from b) is tuned into resonance using nitrogen deposition.}
    
    \label{fig:Gas tuning}
\end{figure*}

\begin{table}[h]
\caption{Yield metrics for diamond QMCs}
\label{tab:yield_metrics}
\begin{tabular}{@{}lll@{}}
\toprule
Stage & Metric & Typical Value \\
\midrule
Mask fabrication     & Pattern yield      & $\sim$100\% \\
Mask transfer        & Transfer yield     & $\sim$80--90\% \\
Chiplet integrity    & Structural yield   & $\sim$75--85\% \\
Cavity formation     & Fill fraction      & $\sim$60\% \\
Spectral alignment   & Tunable yield      & $>$90\% \\
\botrule
\end{tabular}
\end{table}

\section{Analysis of fabrication-induced variation}\label{sec5}
To characterize the fabrication outcome of the processed photonic crystal cavities, we integrated computer vision analysis of SEM images with the measured central wavelength distribution of the fabricated devices (see Methods Section 8.2.1).
We subsequently estimated the thickness distribution of the cavities using a surrogate model trained on FDTD simulations.

\begin{figure*}[t]
    \centering
    \includegraphics[scale=0.74]{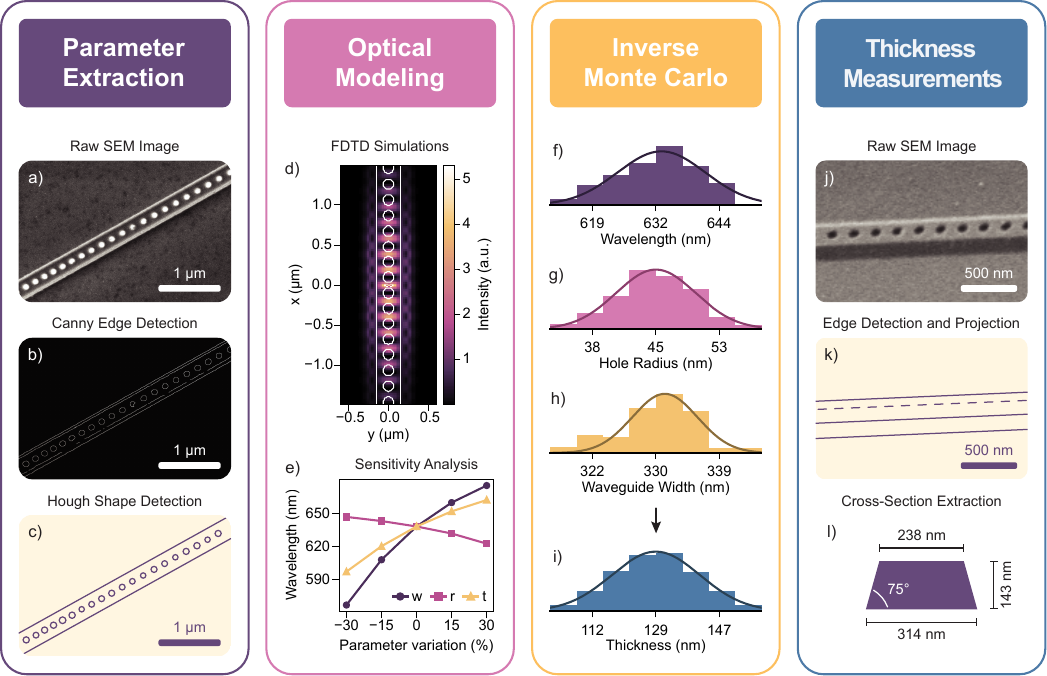}
    \caption{Analysis of fabrication-induced variation.
    \textbf{(a-c)} SEM-based extraction of geometric parameters from fabricated nanobeam cavities.
    \textbf{(d-e)} Sensitivity of cavity resonance wavelength to geometric variations, obtained from FDTD simulations. 
    \textbf{(f-i)} Comparison between experimentally measured wavelength distributions and the range of thickness variation required to account for the observed optical spread.
    \textbf{(j-l)} Independent thickness measurements obtained from tilted SEM images using geometric projection of the waveguide cross-section. Together, these analyses indicate that thickness variation is the dominant contributor to cavity wavelength inhomogeneity and that its magnitude is consistent with direct geometric measurements.
}
    \label{fig:model}
\end{figure*}

Finally, thickness was measured from tilted SEM images using geometric projection of the waveguide cross-section, from which the trapezoidal cross-section geometry was reconstructed and compared with the model predictions. The full workflow is summarized in Fig.~\ref{fig:model}.
We begin by analyzing SEM images from 124 cavities (an example is shown in Fig.~\ref{fig:model}a).
Each image is processed with Canny edge detection (Fig.~\ref{fig:model}b) and subsequently analyzed using Hough shape detection to extract key geometrical parameters, including waveguide width and hole radius, as summarized in Fig.~\ref{fig:model}a-c. Based on the extracted geometries, we perform FDTD simulations to compute the cavity's central wavelength, quality factor, and far-field emission (see Methods Section 8.2.2).
A simulated near-field mode profile is shown in Fig.~\ref{fig:model}d. Representative trends in cavity resonance wavelength as a function of geometric variations are illustrated in Fig.~\ref{fig:model}e.

By combining experimentally measured distributions of cavity resonance wavelengths, waveguide widths, and hole radii (Fig.~\ref{fig:model}f-h) with electromagnetic simulations, we estimated the corresponding cavity thickness distribution, shown in Fig.~\ref{fig:model}i. To assess the accuracy of the constrained analysis, we independently estimated the cavity thickness from angled (45$^\circ$) SEMs (Fig.~\ref{fig:model}j).
We first measured the top and bottom widths from standard vertical SEMs and then applied edge detection to the tilted image (Fig.~\ref{fig:model}k), then we projected the trapezoidal waveguide profile to extract the thickness that best reproduces the observed projection.
The reconstructed cross-section (Fig.~\ref{fig:model}l) yields a thickness value consistent with the distribution obtained from the Monte Carlo inversion, showing consistency between the thickness variation inferred from optical trends and independent geometric measurements.

\section{Discussion and Outlook}\label{sec6}

A central motivation of this work is that our fabrication flow aligns diamond nanophotonics with the broader semiconductor foundry paradigm. Traditionally, diamond nanophotonic devices-especially undercut structures and quantum microchiplets-have relied on small-scale processing in highly specialized laboratories, where electron-beam lithography (EBL) performed directly on the diamond defines device patterns. While this method offers flexibility and high resolution, its serial nature restricts throughput and its sensitivity to process drift limits uniformity and reproducibility. These constraints pose major bottlenecks for scaling from proof-of-concept demonstrations to wafer-scale quantum photonic hardware. In contrast, the method used in this work shifts the patterning from small diamond substrates to wafer-scale silicon produced entirely through a commercial SOI foundry process, leveraging the same process controls and metrology infrastructure that underpin industrial CMOS fabrication. This positions the method as inherently scalable: once the design is taped out, the mask can be reproduced at high volume and high uniformity, independent of local tool availability, operator-dependent EBL optimization, or difficulties arising from fabrication on small insulating substrates. 

While the optical characterization of nanostructures resulting from our process reveal promising results and successful coupling to tin vacancy defects, we can further refine the fabrication to reduce inhomogeneity and achieve improved yield and performance. Our combined optical characterization and modeling reveal key parameters causing variation across the mask, allowing us to identify and compensate for systematic variations. By correlating hyperspectral measurements with geometric parameters extracted from SEM analysis and with trends obtained from a surrogate FDTD model, we identify thickness variation as the primary driver of cavity resonance-wavelength inhomogeneity. Using spatially resolved data taken across the full membrane (Fig. 3e), we map the wavelength distribution resulting from this thickness variation and investigate systematic vs random variations in cavity resonance. We fit the spatial distribution to a surface and examine the resulting residual (Supplementary section 1) using two different methods to capture both long-range and local spatial variations - quadratic surface fitting over the whole mask and a locally estimated scatterplot smoothing (LOESS) fitting method that captures local variations. We find that the residual after the LOESS fitting shows an almost twenty percent reduction compared to the starting distribution of resonances, while the quadratic surface fitting yields a roughly ten percent improvement. This indicates that while the majority of the inhomogeneity is due to random variations in fabrication, local and long-range systemic spatial trends play a role in the distribution as well. For systematic variations occurring routinely run-to-run, we can include intentional variation in the mask design to compensate for these variations, leaving only non-systematic device variations arising from local fabrication imperfections. By further improving our mask transfer methods to prevent wrinkling and tearing of the membrane we can potentially counteract local spatial variations due to mask distortion as well as improve nanostructure yield since the majority of chiplets showing zero successful cavities arose from areas with catastrophic membrane damage.

Finally, the chiplet-based nature of this platform provides robustness to defective chiplets through post-fabrication selection and replacement. If a single integration attempt yields a functional QMC with probability $p$, then allowing up to $r$ replacement attempts reduces the residual defect probability per site to $q^{\,r+1}$, where $q = 1-p$. The expected number of functional QMCs therefore approaches $N_{\mathrm{good}} = N (1 - q^{\,r+1})$, enabling effective yield to be maintained as system size increases. Combined with the sub-linear scaling of integration effort enabled by batch transfer printing, selective replacement decouples chiplet yield from substrate yield and provides a practical pathway toward large-scale quantum photonic systems with bounded integration effort.\\

\section{Methods}\label{sec11}

\subsection{Diamond Nanostructure Fabrication}\label{subsec8.1}

\subsubsection{Diamond Plasma Etching and Undercut}\label{subsubsec8.1.1}

The single crystalline diamond substrate (elementsix ELSC Series, (4.5mm \,$\times$\ 4.5mm \,$\times$\ 0.5mm)), was cleaned prior to stamping with our thin silicon foundry mask by means of a tri-solvent rinse (acetone, methanol, isopropyl alcohol) and a submersion in 3:1 Pirhana solution $(H_\text{2}SO_\text{4}:H_\text{2}O_\text{2})$ for 10 minutes to promote mask surface adhesion and conformal stamping without air gaps stemming from contamination embedded between the substrate and mask.

A vertical anisotropic etch was carried out on the masked diamond substrate using a bias-assisted reactive ion etch (SAMCO RIE-200iP) with plasma species from a pure $O_\text{2}$ carrier gas. 
A conformal coating of 20nm of $Al_{2}O_{3}$ was then deposited by means of Atomic Layer Deposition (Veeco Savannah 200) as a passivation layer to protect the sidewalls of the diamond waveguides defined in the prior anisotropic etch.
Next, a timed bias-assisted  anisotropic reactive ion etch (SAMCO RIE-200iP, Carrier Gas: CF$_{4}$) was employed to remove the passivation Al$_{2}$O$_{3}$ layer from all surfaces on the substrate except those perpindicular to the bias field from the electrostatic chuck, followed by a brief anisotropic etch to create a small lateral plane for the etch to proceed faster, leaving a passivation layer on what will form the sidewalls of the waveguides, while allowing access for the subsequent quasi-isotropic etch $O_{2}$ plasma species to undercut the diamond nanostructures.
Finally, the substrate was quasi-isotropically etched by O$_{2}$ plasma species at an elevated chuck temperature (185\textdegree C), by which the diamond waveguides with embedded color centers are freely suspended from the bulk substrate and left freestanding by strategic artifacts in the mask to serve as a mechanical chassis tethering the nanostructures to the bulk substrate. \cite{Li2015, Wan2020, Wan_Mouradian_2018, MouradianAPL2017}. 

\subsubsection{Mask and Skirt Removal}\label{subsubsec8.1.2}

Following the conclusion of processing and SEM verification of structure cross-section, the substrate is submerged in Hydrofluoric acid (49\% w/v) for 10 minutes to dissolve the remainder of the Al2O3 ALD sidewall passivation layer. 
Following decontamination, the substrate is further submerged for 12 hours in a heated KOH solution (30\% w/v, 95\textdegree C) to remove the silicon hardmask. The substrate lastly undergoes a critical point drying cycle in IPA to avoid surface tension at the liquid-vapor boundary potentially straining and collapsing the suspended nanostructures. 

\subsection{Thickness estimation and simulation support}\label{subsec8.2}

Geometric parameters are extracted from SEM images using a Python pipeline (OpenCV, scikit-image): top-view SEMs measure waveguide width and hole radius; tilted SEMs provide cavity thickness and sidewall angle. The full implementation, configuration files, extended analysis scripts, and derivation notebooks are available at \url{https://github.com/QPG-MIT/Photonic-Crystal-Cavity-Simulations}.

\subsubsection{Preprocessing and calibration}

Raw SEM images are converted to 8-bit grayscale and denoised using a bilateral filter followed by non-local means denoising to suppress noise while preserving edge sharpness. A Canny edge detector with adaptive thresholds optimized for diamond imaging at 3\,kV provides a stable edge map. Images are cropped to the cavity region, and the pixel scale (nm/px) is obtained from the SEM scale bar.

\subsubsection{Top-view width and hole-radius extraction}

For top-view metrology, we estimate the global nanobeam axis using the structure tensor computed from Scharr gradients, which defines a normal direction across the beam. Signed intensity gradients along this normal produce histogram peaks corresponding to the four edges of the trapezoidal cross-section; the outermost opposite-polarity peaks define the bottom width. Edge positions are refined using iteratively reweighted least squares with weights proportional to the local gradient magnitude. Circular holes in the central strip between the rails are detected using a combination of Hough-circle and contour-based fitting, then filtered by radius range, distance from the midline, and minimum spacing along the beam to yield the hole-radius distribution.

\subsubsection{Thickness extraction from tilted SEMs}

We analyze \(45^\circ\) tilted SEMs of the trapezoidal nanobeam (top width \(W_t\), bottom width \(W_b\), thickness \(t\)) to obtain an independent thickness estimate. Under tilt, three horizontal ridges are visible: the far and near top edges and the near bottom edge. After preprocessing (contrast-limited adaptive histogram equalization, hole filling, Canny edge detection), these ridges are detected using a probabilistic Hough transform and fit with robust lines using a Welsch loss function. From the fitted lines we measure two perpendicular separations along the image normal: \(d_T\) (between far and near top ridges) and \(d_{N\!B}\) (between near top and near bottom ridges). Together with \(W_t\) and \(W_b\) from the top-view analysis, the known stage tilt angle \(\theta\), and the apparent image angle \(\psi\) of the nanobeam, the projection geometry yields
\begin{equation}
  t = \frac{\cos^2\theta\,\cos\psi}{\sin\theta\,\sqrt{1 - \sin^2\theta\,\cos^2\psi}}
      \left(\frac{W_t\,d_{N\!B}}{d_T} - \frac{W_b - W_t}{2}\right).
\end{equation}
This scale-free form depends only on the ratio \(d_{N\!B}/d_T\), not the absolute pixel size. The full derivation is provided in the GitHub repository. Uncertainties in \(W_t\), \(W_b\), \(d_T\), and \(d_{N\!B}\) are propagated to \(t\) via Monte Carlo sampling: widths are drawn from their measured distributions, separations are perturbed according to estimated pixel and line-fit noise, and \(t\) is recomputed using the formula above.

\subsubsection{FDTD Simulations}\label{subsubsec8.2.2}

Finite-difference time-domain (FDTD) simulations were performed using Tidy3D, a cloud-accelerated electromagnetic solver. Diamond optical properties are modeled using a standard two-term Sellmeier dispersion
with coefficients $B_1 = 0.3306$, $C_1 = (0.175\,\mu\text{m})^2$, $B_2 = 4.3356$, $C_2 = (0.106\,\mu\text{m})^2$, valid for wavelengths $0.23$--$5\,\mu$m. At the SnV emission wavelength of $\lambda = 620$\,nm, this yields $n = 2.424$. Perfectly matched layer (PML) boundary conditions are applied on all six faces to absorb outgoing waves and prevent spurious reflections from domain boundaries. The excitation source is a point dipole with $E_y$ polarization positioned at the cavity center, driven by a Gaussian pulse with center frequency corresponding to $\lambda_0 = 620$\,nm and relative bandwidth $\Delta f/f_0 = 12\%$. Cavity quality factors are extracted from time-domain field ringdown data using the Tidy3D ResonanceFinder plugin based on Prony's method. A point field monitor records $E_y(t)$ at the cavity center with temporal resolution of 5 simulation time steps.


\subsubsection{Inverse Monte Carlo}\label{subsubsec8.2.3}

Cavity thickness is inferred from measured resonance wavelengths using a Monte Carlo inversion based on a surrogate model. The stochastic input parameters and their sampling models are summarized in Table~\ref{tab:imc_inputs}.

\begin{table}[h]
\centering
\caption{Input parameters for inverse Monte Carlo thickness estimation.}
\label{tab:imc_inputs}
\begin{tabular}{llll}
\hline
Parameter & Symbol & Source of distribution & Sampling model \\
\hline
Waveguide width & $W$ & Top-view SEM & $\mathcal{N}(\bar{W},\sigma_W)$ \\
Hole radius & $r$ & Top-view SEM & $\mathcal{N}(\bar{r},\sigma_r)$ \\
Resonance wavelength & $\lambda$ & Hyperspectral measurement & Empirical distribution \\
\hline
\end{tabular}
\end{table}

\noindent The empirical wavelength distribution is drawn directly from the measured resonance dataset. For each sampled triplet $(W_j, r_j, \lambda_j)$, the corresponding thickness $t_j$ is obtained by solving
\begin{equation}
f(W_j,r_j,t_j) = \lambda_j,
\end{equation}

\medskip
 
Repeating the inversion over $N_{\mathrm{MC}}$ samples yields an ensemble of thickness values $\{t_j\}$ consistent with the measured geometric and optical variability. 
The inferred thickness distribution is characterized by its mean and standard deviation:
\begin{eqnarray}
\bar{t} &=& \frac{1}{N_{\mathrm{MC}}} \sum_{j=1}^{N_{\mathrm{MC}}} t_j, \\
\sigma_t &=& \sqrt{\frac{1}{N_{\mathrm{MC}}-1} \sum_{j=1}^{N_{\mathrm{MC}}} (t_j - \bar{t})^2}.
\end{eqnarray}

The inverse Monte Carlo analysis was performed using $N_{\mathrm{MC}} = 10{,}000$ samples with input parameters $W = 0.330 \pm 0.004$\,\textmu m, $r = 0.045 \pm 0.004$\,\textmu m, and $\lambda = 633.2 \pm 9.0$\,nm. The resulting thickness distribution has mean $\bar{t} = 0.129$\,\textmu m and standard deviation $\sigma_t = 0.012$\,\textmu m.

\subsection{Fabrication spatial analysis}\label{subsec8.3}



Given the spatially resolved cavity data in Figure 2, we can analyze spatial trends in optical properties across the mask and use this to inform and improve our future fabrication. We begin by fitting polynomial surfaces to the cavity resonance wavelength data from Figure 2e, taking each cavity in the center of a nanobeam as a point in our grid of data. After fitting the data using a least squares method and a local polynomial regression (LOESS) method, we analyze the resulting histogram of values before and after subtracting the fitted surface from the data. Finally for each fit, we perform a semivariogram residual analysis to examine how each fitting method performs in terms of removing systematic spatial variations from the data. By examining the spatial trends across the mask, their length scales, and magnitude vs random variations, we can potentially counter any systematic trends due to fabrication processes and improve the homogeneity of our devices in future fabrication runs.

\subsubsection{Surface fitting}
We plot the difference between the center frequency of each cavity and the target wavelength (620 nm) and use a least squares fitting method to fit a quadratic polynomial surface to the resulting distribution. Supplementary figure 2a shows the best fit quadratic surface overlaid with the data points in a 3D plot. Subtracting the surface fit from each data point we are left with the residual values, letting us compare the inhomogeneous distribution of resonance wavelengths before and after removing systematic variations (Supplementary figure 2b).

We observe a small reduction in the width of the distribution, reflected in a decrease in standard deviation of 10.3 percent, from 8.44 nm to 7.57 nm.  In order to capture more local systematic spatial variations, we next employ a local polynomial fitting technique - local regression, or LOESS. In this method we fit a locally weighted linear model at each data point. For each point, we identify the k nearest neighbors, and compute their average distance to the target point, $u_{ij}$, where distance $d_{ij}$ is defined as
\begin{equation}
d_{ij} = \sqrt{(x_j - x_i)^2 + (y_j - y_i)^2}.
\end{equation}
\begin{equation}
u_{ij} = \frac{d_{ij}}{d_{\max}},
\end{equation}

We then apply a tricube kernel to weight neighbors by normalized distance:
\begin{equation}
w_{ij} =
\begin{cases}
\left(1 - u_{ij}^3\right)^3, & u_{ij} < 1, \\
0, & u_{ij} \ge 1.
\end{cases}
\end{equation}

We use this set of weighted points to fit a plane defined by 
\begin{equation}
z \approx a + b x + c y,
\end{equation}

and the fitted local model is evaluated for each point in the grid to obtain a smooth estimate of the spatial trend (Supplementary Figure 3a).

We observe a larger reduction in the width of the distribution with the standard deviation decreasing by about 19 percent, from 8.44 nm to 6.84 nm (Supplementary Figure 3b). We ascribe the improved performance of the local fitting compared to the single quadratic surface to the presence of local fabrication variations that cannot successfully be captured by a single surface. This is likely due to a combination of deformations to the Si mask membrane during stamping (visible in optical microscope images of the fabricated sample), as well as short-range fabrication variations due to proximity effect or mask design.

\subsubsection{Residual analysis}

Finally, we analyze the residuals after both quadratic and LOESS fits compared to the original data to validate how successfully each model removed systematic fabrication variations. We compute the semivariogram of the raw and residual data - the average squared difference between pairs of data points (semivariance) as a function of distance between the points. Due to differences in the absolute distance between the X coordinate and Y coordinate of our spatially mapped cavity data, we examine the residuals for each axis separately. Supplementary Figure 4 panels a) and b) show results for the raw data and LOESS fits, while panels c) and d) show results for the quadratic surface. For data with no spatial trends, the semivariogram should be flat as a function of distance. Local random noise shows up as a "nugget" near zero, while spatial correlations show up as an increase in the semivariance as a function of distance. The raw values (blue dots and trendline) show a consistent increase as a function of distance. For both the X and Y data, the LOESS residual data (orange points and line) remains close to flat after an initial increase. This initial increase is likely due to shortrange spatial correlations not captured by our LOESS fitting. The quadratic surface removes significantly less of the spatial correlations, consistent with the smaller reduction in inhomogeneous distribution of cavity resonances for the residual after quadratic fitting vs the LOESS model.


\section*{Acknowledgements}

The authors thank Hamza Raniwala for reviewing the manuscript and providing valuable feedback. J.A. acknowledges the support from KACST-MIT Ibn Khaldun Fellowship for Saudi Arabian Women at MIT, from Ibn Rushd Postdoctoral Award from King Abdullah University of Science and Technology (KAUST), and from the Army Research Office MURI (Ab-Initio Solid-State Quantum Materials) Grant No. W911NF-18-1-043. This work was partially supported by the MITRE Quantum Moonshot Project.

\section*{Data availability}

Correspondence and requests for materials should be addressed to the correspondig authors.

\section*{Author contribution}

D.E., J.A., and L.L.  conceived the project. J.A. and L.L. developed the mask design. M.S. characterize the diamond plate. A.K. carried out the diamond fabrication with inputs from J.A. and G.C. J.A. and A.B. etched, suspended and stamped the masks. G.C. performed the optical characterization and gas tuning. W.Y. wrote the optical scanning codes. A.B. performed the thickness estimation and simulation support. G.C. performed the fabrication spatial analysis. J.A., A.B. and G.C. wrote the manuscript. All authors reviewed and approved the manuscript.

\section*{Competing interests}

D.E. and J.A. are co-founders of a company developing and commercializing diamond quantum photonic devices. The remaining authors declare no competing interests.





\bibliography{sn-bibliography}

\end{document}


\title{Foundry-Enabled Patterning of Diamond Quantum Microchiplets for Scalable Quantum Photonics}
\maketitle


\begin{figure}[]
\centering
\includegraphics[width=1\textwidth]{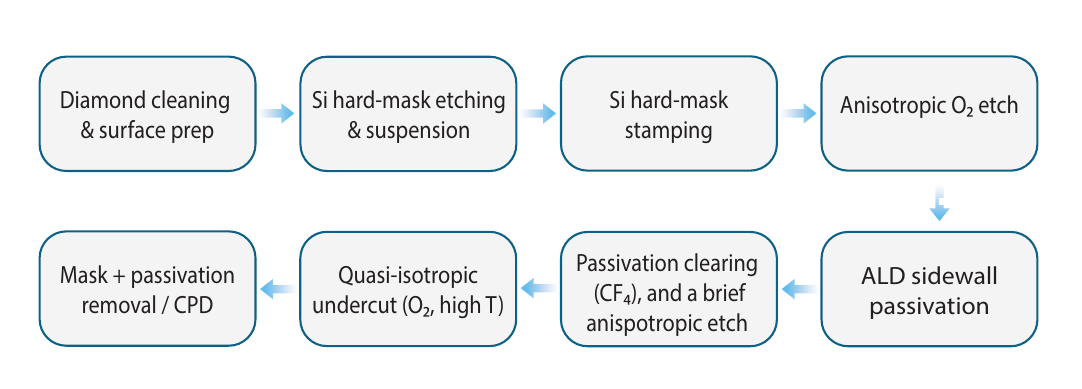}
\caption{Process flow for diamond nanostructure fabrication using transferred silicon hard masks.}
\label{Extended Data Fig:signed_gradient}
\end{figure}

\begin{figure}[h]
\centering
\includegraphics[width=1\textwidth]{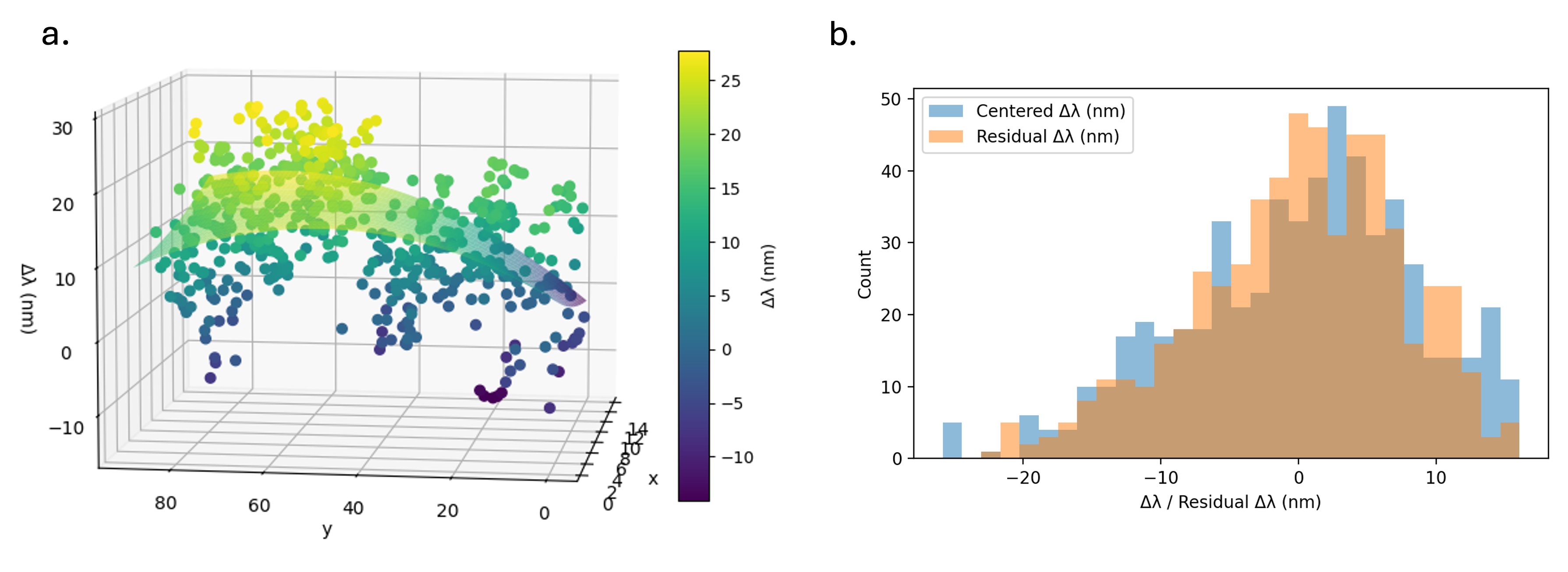}
\caption{a) Quadratic surface fitted to cavity resonance wavelength data. X and Y are the index of each cavity wavelength data point within the mask. b) Histograms showing difference between measured cavity wavelength and target wavelength (620 nm) with and without the fitted surface removed. }
\label{fig:Quadratic_fit}
\end{figure}

\begin{figure}[h]
\centering
\includegraphics[width=1\textwidth]{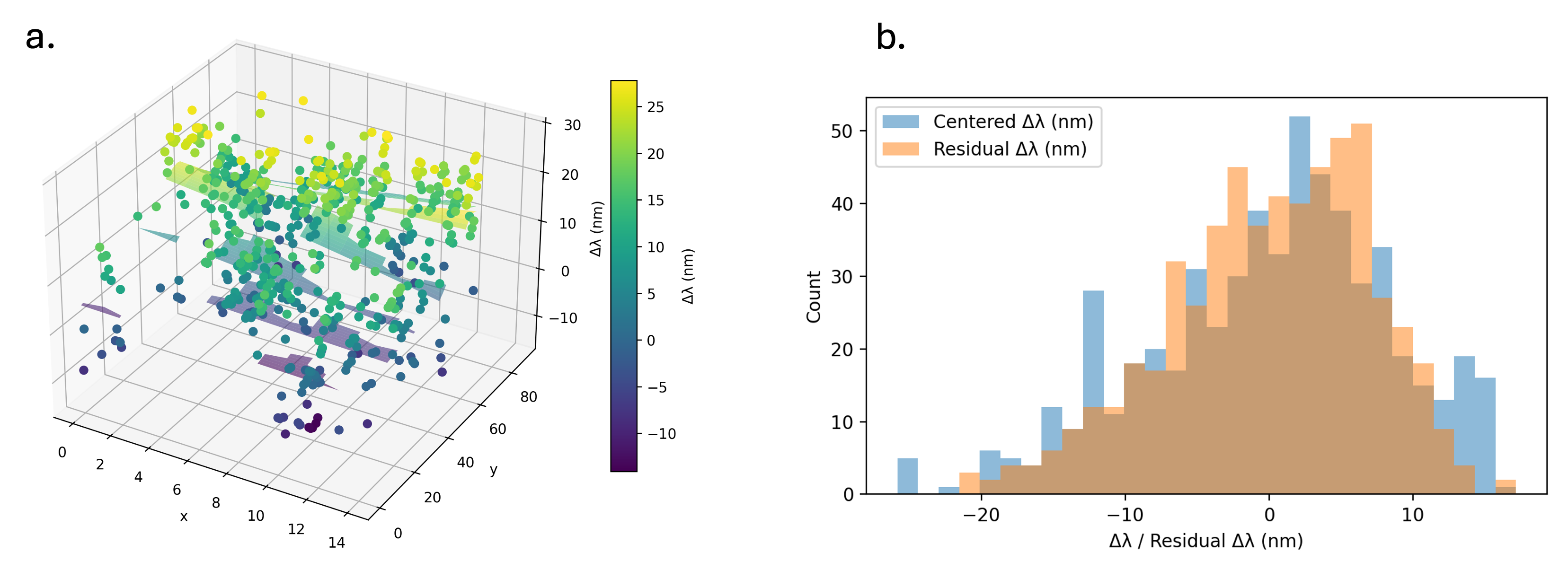}
\caption{a) Rendition of the planar surfaces fitted using the LOESS method. Not all surfaces are shown.  b) Histograms showing difference between measured cavity wavelength and target wavelength (620 nm) with and without the LOESS fit. }
\label{fig:LOESS_fit}
\end{figure}

\begin{figure}[h]
\centering
\includegraphics[width=1\textwidth]{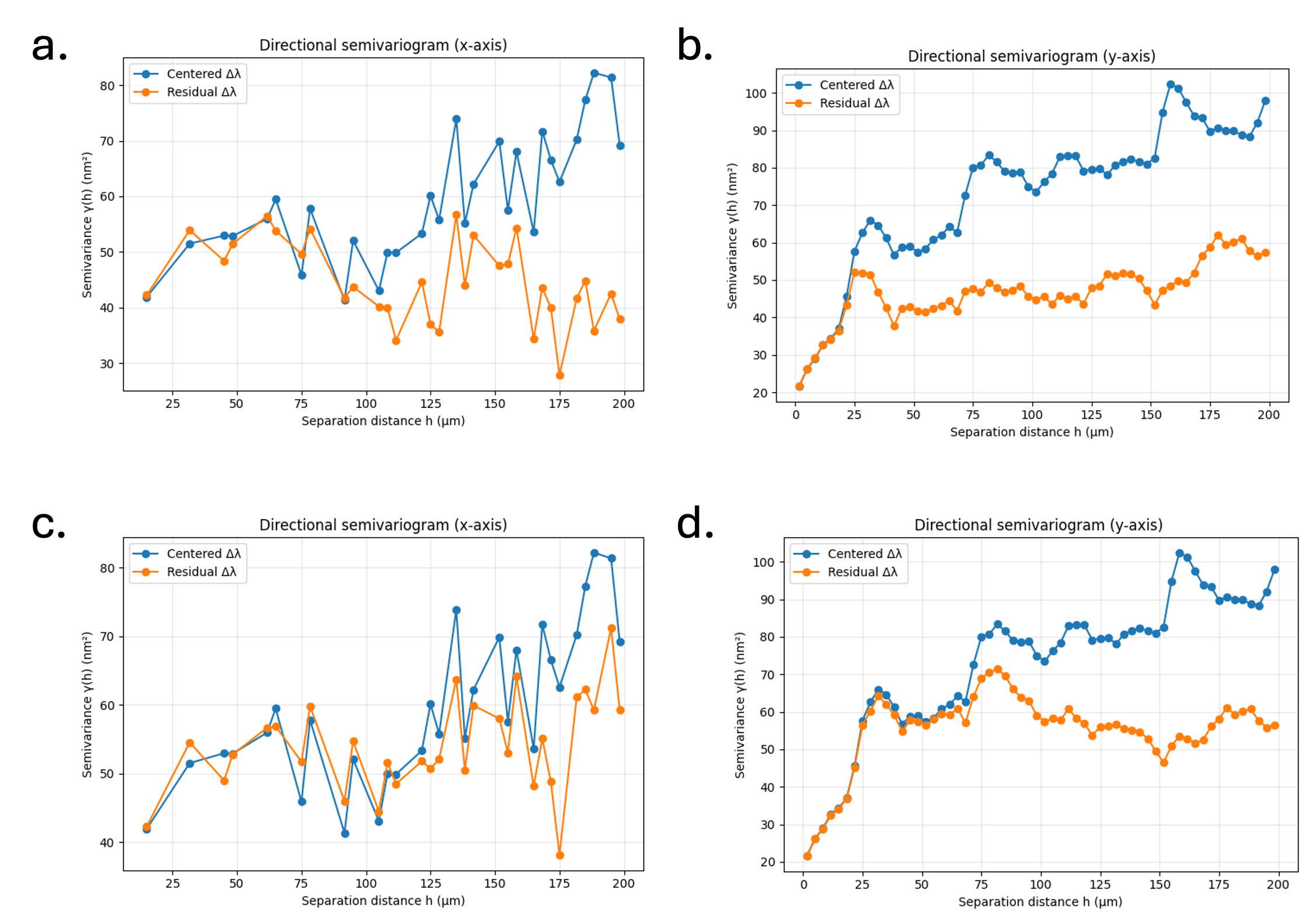}
\caption{a) Semivariogram along the X data axis for raw data vs residual after LOESS fitting. X axis is in microns corresponding to real separation distance in the mask. b) Semivariogram along the Y data axis for raw data vs residual after LOESS fitting. Y axis is in microns corresponding to real separation distance in the mask. c) and d) The same semivariogram for raw data vs residual after quadratic fitting. }
\label{fig:Residual_comparison}
\end{figure}

\begin{figure}[h]
\centering
\includegraphics[width=1\textwidth]{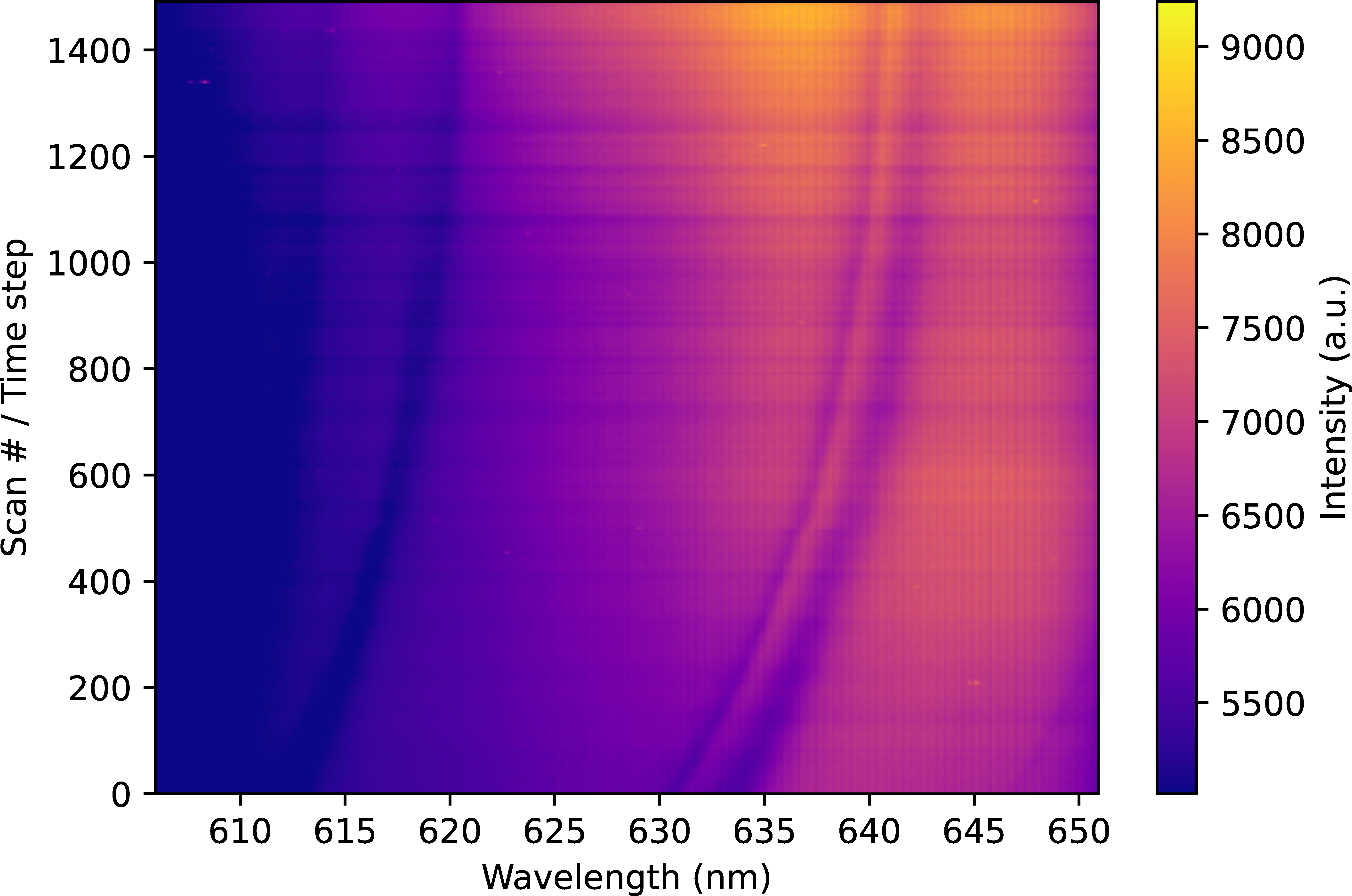}
\caption{a) Cavity reflection measurements as a function of time as nitrogen gas is introduced into the cryostat chamber. The cavity with a resonance near 633 nm before tuning shows more than 9 nm of redshift as gas is deposited.}
\label{fig:Residual_comparison}
\end{figure}